\documentclass[manuscript]{aastex61}

\newcommand{\psj}{Planet. Sci. J.} 

\received{7 April 2022}
\revised{2 November 2023}
\accepted{2 November 2023}
\submitjournal{The Astronomical Journal}

\shorttitle{NEOWISE Observations of Distant Active Comets B1, K2, \& U3}
\shortauthors{Milewski et al. 2024}
\usepackage{graphicx}
\graphicspath{ {images/} }
\usepackage{gensymb}
\usepackage{lineno}

\begin{document}


\title{NEOWISE Observations of Distant Active Long-period Comets C/2014 B1 (Schwartz), C/2017 K2 (Pan-STARRS), and C/2010 U3 (Boattini)}


\correspondingauthor{Dave G. Milewski}
\email{dgmilewski@epss.ucla.edu}

\author[0000-0002-7841-9464]{Dave G. Milewski}
\affil{Department of Earth, Planetary, and Space Sciences, University of California at Los Angeles, \\
595 Charles E. Young Drive East, \\
Los Angeles, CA 90095-1567, USA}

\author[0000-0003-2638-720X]{Joseph R. Masiero}
\affil{Caltech/IPAC, \\
1200 E. California Boulevard, MC 100-22, \\
Pasadena, CA 91125, USA} 

\author[0000-0002-5736-1857]{Jana Pittichová}
\affil{NASA Jet Propulsion Laboratory/California Institute of Technology, \\
4800 Oak Grove Drive, \\
Pasadena, CA 91109-8001, USA}

\author[0000-0003-0457-2519]{Emily A. Kramer}
\affil{NASA Jet Propulsion Laboratory/California Institute of Technology, \\
4800 Oak Grove Drive, \\
Pasadena, CA 91109-8001, USA}

\author[0000-0002-7578-3885]{Amy K. Mainzer}
\affil{University of Arizona, \\
1629 E. University Boulevard, \\
Tucson, AZ 85721, USA}

\author[0000-0001-9542-0953]{James M. Bauer}
\affil{Department of Astronomy, University of Maryland, College Park, \\
College Park, MD 20742-2421, USA}


\begin{abstract}
Hyperactive comet activity typically becomes evident beyond the frost line ($\sim$3–4 au) where it becomes too cold for water-ice to sublimate. If carbon monoxide (CO) and carbon dioxide (CO\textsubscript{2}) are the species that drive activity at sufficiently large distances, then detailed studies on the production rates of these species are extremely valuable to examine the formation of the solar system because these two species (beyond water) are next culpable for driving cometary activity. The NEOWISE reactivated mission operates at two imaging bandpasses, $W1$ and $W2$ at 3.4~\micron~and 4.6~\micron, respectively, with the $W2$ channel being fully capable of detecting CO and CO\textsubscript{2} at 4.67~\micron~and 4.23~\micron~in the same bandpass. It is extremely difficult to study CO\textsubscript{2} from the ground due to contamination in Earth's atmosphere. We present our $W1$ and $W2$ photometry, dust measurements, and findings for comets C/2014 B1 (Schwartz), C/2017 K2 (Pan-STARRS), and C/2010 U3 (Boattini), hereafter, B1, K2, and U3, respectively. Our results assess CO and CO\textsubscript{2}~gas production rates observed by NEOWISE. We have determined: (1) comets B1 and K2 have CO\textsubscript{2}~and CO gas production rates of $\sim$10\textsuperscript{27} and $\sim$10\textsuperscript{29} molecules s\textsuperscript{–1}, respectively, if one assumes the excess emission is attributed to either all CO or all CO\textsubscript{2}; (2) B1 and K2 are considered hyperactive in that their measured $Af\rho$ dust production values are on the order of $\gtrsim$10\textsuperscript{3} cm; and (3) the CO and CO\textsubscript{2} production rates do not always follow the expected convention of increasing with decreased heliocentric distance, while B1 and K2 exhibit noticeable dust activity on their inbound leg orbits.
\end{abstract}




\section{Introduction} \label{sec:introd}
Comets are 4.6-billion-year-old surviving relics of the Sun’s protoplanetary disk and, in addition to water, contain volatile inventories of carbon monoxide (CO) and carbon dioxide (CO\textsubscript{2}). The two major source regions for comets are thought to be the Oort Cloud (Oort 1950) and Kuiper Belt (Jewitt \& Luu 1993) at temperatures of $\sim$10 K and $\sim$40 K, respectively. The most famous model for cometary mass loss is based on Whipple (1950), where mass loss results from the sublimation of water ice from a central nucleus. However, beyond 5 au, other processes must be responsible for cometary activity because temperatures are too low for sublimation of water ice, as seen with populations of outer solar system bodies such as the Centaurs (Wierzchos et al. 2017; Womack et al. 2017; Lisse et al. 2022; Bodewits et al. 2020). CO is difficult and, in the case of CO\textsubscript{2} extremely difficult (given the observation technique), to study through the Earth’s CO\textsubscript{2}-rich atmosphere. This makes space-based observatories especially useful in studies of CO and CO\textsubscript{2} species in comets by enabling observing outside of the Earth’s atmosphere.\newline

The Wide-Field Infrared Survey Explorer (hereafter WISE) spacecraft was launched on UT 2009 December 14, with the goal of conducting a full infrared all-sky survey in the thermal bands 3.4, 4.6, 12, and 22~\micron~(hereafter $W1$, $W2$, $W3$, and $W4$; Wright et al. 2010) in the mid-infrared spectrum, building on that of Infrared Astronomical Satellite. After its cryogens were depleted, the WISE spacecraft was put into hibernation in 2011. It was later reactivated on UT 2013 December 13 and renamed as the Near-Earth Object Wide-field Infrared Survey Explorer (NEOWISE) because it continued a 2-band infrared survey (Mainzer et al. 2011). The goal of the NEOWISE Reactivation Mission is to find and characterize potential solar system small bodies such as asteroids and comets that may pose a threat to the Earth under NASA's strategic goal for planetary defense. The NEOWISE Reactivated Mission differs from the WISE-prime mission by using only the $W1$ and $W2$ channels and being passively cooled to approximately 75 K, allowing a continuation of an infrared all-sky survey for solar system small bodies. There are a total of 274 comets as of 2023 September that have been detected by the NEOWISE spacecraft, with those comets sometimes having multiple observations. Just half the number of the population of NEOWISE observed comets consist of short-period comets (SPCs), while the other half are considered to be long period comets (LPCs); (Bauer et al. 2017). SPCs have orbital periods of $\lesssim$200 yr, while the LPC orbital periods often exceed 200 yr (Vokrouhlick{\'y} et al. 2019). At 3.4 \micron, the $W1$ channel is sensitive to sunlight reflected from cometary dust particles when they are close to the Sun, as well as thermal emission from them. The $W2$ channel at 4.6 \micron~is also sensitive to reflected sunlight and thermal emission from dust, but emission from the 4.67 and 4.23~\micron~bands of CO and CO\textsubscript{2} are also detectable. These peak wavelengths refer to the fundamental CO $\nu$ = 1–0 and CO\textsubscript{2} $\nu_3$ emission lines, respectively (Mainzer et al. 2014; Bauer et al. 2015). In this work, we compute the combined production rates for CO and/or CO\textsubscript{2} as well as dust production rates (by way of $Af\rho$ calculations) for comets C/2014 B1 (Schwartz), C/2017 K2 (Pan-STARRS), and C/2010 U3 (Boattini), hereafter, B1, K2, and U3, respectively, and discuss the significance of our findings.\newline

\subsection{Comets C/2014 B1 (Schwartz), C/2017 K2 (Pan-STARRS), and C/2010 U3 (Boattini)} \label{sec:cometsbackground}
B1 was discovered on UT 2014 January 28 (Schwartz \& Sato 2014) at a heliocentric distance $r_H$~$\sim$~12 au. Optical images of B1 show a complex morphology that resembles a biconvex lens or \textquotedblleft{discus}\textquotedblright~seen edge-on, which has been explained to be due to ejection of material from its equatorial region (Jewitt et al. 2019a). Perihelion passage of B1 occurred on UT 2017 September 10 at a distance $q$ = 9.557 au, just around the orbit of Saturn.\newline

Discovered on UT 2017 May 21 (Wainscoat et al. 2017) at $r_H$ = 16.1 au, K2 was the previous record holder for the furthest inbound active comet at $r_H$ = 23.7 au found in optical precovery data (Hui et al. 2018). It reached perihelion on UT 2022 December 19 at $q$ = 1.798 au. Numerous studies of K2, e.g., Jewitt et al. (2017, 2019b, 2021), Meech et al. (2017), Hui et al. (2018), and Yang et al. (2021), have suggested that CO is the likely driver of activity. K2 was found to be active at $r_H$ = 23.7 au, only to be recently bested by the previous furthest-ever inbound active comet, U3, at $r_H$ = 25.8 au. However, in this work, we explore the possibility that CO\textsubscript{2} may also play a role in driving the observed activity.\newline

Finally, U3 was discovered on UT 2010 October 31 at $r_H$ = 18.3 au (Boattini et al. 2010). U3 was previously the furthest inbound comet found to be exhibiting activity at distance $r_H$ = 25.8 au from precovery data (Hui et al. 2019). Perihelion occurred on UT 2019 February 26 at $q$ = 8.446 au. In 2021, this record was overtaken by the discovery of activity on comet C/2014 UN271 (Bernardinelli–Bernstein); (Bernardinelli et al. 2021). C/2014 UN271 was found to be active in the Dark Energy Survey data at a current record distance of $r_H$ = 29 au. With regards to U3, one or a combination of CO and/or CO\textsubscript{2} volatiles may drive activity for this comet. However, few studies have been performed on this object because it is now outbound. A list containing orbital elements for B1, K2, and U3 can be found in Table 1.\newline

All three of these comets are important to help us understand why these objects exhibit such strong activity at such large heliocentric distances. Assessing their composition for supervolatile species is important to understand how much of their primordial inventories still remain, especially for comets active at such large $r_H$.\newline


\section{Observations and Data Collection} \label{sec:obsdata}
We used data taken by WISE (Wright et al. 2010) and NEOWISE (Mainzer et al. 2014). NEOWISE is in a sun-synchronous polar orbit with an orbital period of $\sim$95 minutes, which completes a full mapping of the sky in the infrared twice a year. The satellite's orbit is such that it is facing away from the Sun, usually around solar elongation angles of $\sim$90$^{\circ}$–110$^{\circ}$ (due to orbital precession) and is protected with a sun-shield to prevent stray light from entering the aperture. The telescope aperture is 40 cm, it has a 47\arcmin~$\times$~47\arcmin~field of view, a pixel scale of 2.\arcsec75 pixel\textsuperscript{–1}, and the images have a Full-Width at Half Maximum (FWHM) of 6.\arcsec1~and 6.\arcsec4~in $W1$ and $W2$ respectively. Data in $W1$ and $W2$ are taken simultaneously every 11 s by use of a dichroic beamsplitter to separate the two channels. Each individual frame has an exposure time of 7.7 s in both $W1$ and $W2$. Because of its rate of motion on the sky, there are overlaps in data coverage, which means that moving objects such as solar system small bodies can be positively identified. Initial data products have instrumental and astrometric corrections applied via the WISE Science Data Center (Cutri et al. 2012). Once the calibrated images are searched for moving object detections, their positions and times are reported to the Minor Planet Center several times per week (Cutri et al. 2015). The calibrated image and source detection data are released online publicly via the Infrared Processing and Analysis Center (IPAC) Infrared Science Archive (IRSA) (see: https://irsa.ipac.caltech.edu/). The NEOWISE spacecraft takes data for many solar system objects over a variety of different heliocentric distances and observation epochs. Images are collected every $\sim$3 hr for most moving objects. Due to the typical rate of motion of most minor planets in the inner solar system through the NEOWISE scan circle, the \textquotedblleft{visit}\textquotedblright~length usually spans $\sim$36 hr. Objects are sometimes detected on multiple epochs, and we define a \textquotedblleft{visit}\textquotedblright~as a set of observations separated in time from one another by approximately a week or more~(Bauer et al. 2015). \newline

\section{Methods and Aperture Photometry} \label{sec:methodsandphot}
\subsection{Methods} \label{sec:methods}
After collecting and grouping all the available data corresponding to a particular visit for a comet, we sorted the individual single-exposure frames by the respective $W1$ and $W2$ bands. From the data collected in single frames, we then visually inspected each image to check for detection of the comet at the predicted position from the JPL HORIZONS Web Interface. Our selection criteria for removing frames from further consideration in our study were based on any frame with noticeable aberrations, such as those where there are obvious smearing effects from trailing, over-saturation from the Moon or other bright field objects, an object of interest being cut off by being located at the edge of a frame, or if a particular frame was listed in the IRSA WISE Image Search having a quality assessment of less than 5~(e.g., with the flag $qual frame$ $\textless$5, with 0 being the lowest and 10 the highest) (Cutri et al. 2015). The frames containing useful data were then input into the Image Co-addition with Optional Resolution Enhancement package (hereafter, ICORE; Masci, 2013), to co-add the single-exposure images in the moving reference frame of the object according to predicted positions by JPL HORIZONS. The resulting frames are stacked to position the comet at the very center of each frame, as seen in the Figure 1 dataset showing the morphology of specific comet visits which coincide with the data listed in Table 2. ICORE is an image co-addition package that produces final mosaicked co-adds that matches backgrounds, contains pixel outlier rejection, and boosts the signal-to-noise ratio (S/N) of the individual frames. The final co-added images have a resampled pixel scale of 1\arcsec~pixel\textsuperscript{–1}.\newline

Figure 1 contains the respective $W1$ and $W2$ composite stacked frames, as well as combined $W1$+$W2$ color-mapped composite stacked frames for the specified visits, while Table 2 contains a summary of the data and measurements. Because of the slow rate of motion of the comets on the sky, in some instances there are some visible streaks and artifacts from background stars and objects not being fully suppressed in the final co-added images. The documentation in the ICORE package by Masci et al. (2013) suggests that it does its best to remove any stars bright enough to contaminate sources against moving objects, such as those from the solar system. However, there are still potential low brightness sources, such as galaxies, which may remain in the field. In this case, we refer to Mainzer et al. (2014). Systematic noise in the background may play a factor in driving up the S/N for very faint objects in the background. However, our work does show that the NEOWISE spacecraft is fully capable of detecting objects at these large distances, although it may not be fully possible to remove all sources from the background. Either way, as the object moves, using the NEOWISE data, we opt to exclude frames where very bright stars within the photometry aperture could potentially contaminate the photometry. As in previous work, we adopt the convention of Bauer et al. (2015, 2017) and Stevenson et al. (2015) with $W1$ mapped to blue and green, and $W2$ mapped to red. All frames are rotated such that the orientation always has North up, East to the left, and the size of the image was set at $\sim$4.\arcmin5~$\times$~$\sim$4.\arcmin5. The contrast and stretch are not the same in the images to suppress some of the remaining artifacts. In most cases (with the exception of some visits involving K2), the comets cannot be seen in the individual exposures but do, however, appear in the co-added images as point sources, sometimes with extended features suggestive of dust or comae. Table 2 also contains a column that lists what we interpret as a comet visual detection or non-detection for the $W1$ and $W2$ channels from the images.\newline

\subsection{Aperture Photometry} \label{sec:aperturephot}
Aperture photometry was performed on individual objects using a modified version of the publicly available Astropy ‘photutils’ package. The photometry was executed for known objects with magnitudes and errors presented in the NEOWISE Reactivation Catalog Data (Cutri et al. 2012). We used magnitude zero-points that are specific for $W1$ and $W2$ (Cutri et al. 2015), and then performed the aperture photometry on the ICORE frames. Counts were converted to magnitudes, and then to flux densities with the correct band specific zero-points listed in Wright et al. (2010). Aperture photometry was also performed on point sources in the single-exposure frames that were used to make the final co-added images to ensure that the field stars were of consistent brightness reported in the NEOWISE Single-Exposure Point Source Catalog. A variety of apertures were used and it was found that the 11\arcsec~radius aperture was the best choice because it captured nearly all of the light of the source while leaving a small amount of sky background necessary for removal in a sky annulus. Background subtraction from the median value in a sky annulus extending from 50\arcsec~to 70\arcsec~was used to remove sky background. After comparing the results of our single-exposure point source NEOWISE aperture photometry measurements with the values listed in the catalog, we found small deviations on the order of a few percent, which happens because the catalog uses point-spread function (PSF)-fit photometry. This gives us confidence in performing aperture photometry on the stacked images. Consistent with the results of Bauer et al. (2015, 2017), Stevenson et al. (2015), and Rosser et al. (2018), the 11\arcsec~aperture was also chosen because it best captures the FWHM of the PSF for NEOWISE observations and encloses most of the light for the target with little to no significant change in magnitude for various similar size apertures while avoiding nearby field objects. Our measurement technique was also validated against several distant solar system small bodies. Because of the distance and faintness of the comets that we were measuring, the major sources of photometric error are the result of the absolute calibration uncertainty and systematics, which are often due to faint inertial object residuals remaining in the final co-adds. On most visits (with the exception of K2's passage into the frost line around $\sim$3.5 au), because the phase angle was very small (see Table 2), we elected to ignore phase angle corrections in our measured flux densities. The dominant source of flux in $W1$ at this distance appears to be the result of reflected light of dust grains in the coma which are of unknown size, porosity, and shape. Figure 1 shows that most objects are clearly visible, diffuse, and exhibit activity (sometimes observed in individual exposures) not typically seen for comets observed in the NEOWISE data when compared to a stellar point source. Many ‘active’ comets are typically first seen starting around $\sim$5–6 au in NEOWISE data. However, in this work we found that for the case of K2, it was clearly visible and active as far as $\sim$8–9 au in both the $W1$ and $W2$ bandpasses, hence deserving merit of future study.\newline

\section{Results and Discussion} \label{sec:resultsdiscussion}
\subsection{Dust Production Rates} \label{sec:dust}
As previously mentioned, the $W2$ channel contains reflected sunlight and some thermal emission. However, at the distances of the times of the observations, the blackbody temperatures of these bodies are too cold to show significant thermal emission in $W2$, making the reflected light signal the dominant source of flux. Because of this, we need to extrapolate the expected $W2$ contribution using the $W1$ flux by creating a reflected light and thermal model to determine how much, if any, excess flux in $W2$ exists that is assumed to be attributed to CO and $\rm CO_2$ gas emission. The $W1$ signal contains reflected light from dust grains and is quantified by the flux measured at 3.4~\micron. Using this flux, we convert it to $Af\rho$ which is a proxy for dust production measurements introduced in A'Hearn et al. (1984). $Af\rho$ stands for Albedo, $A$, filling factor, $f$ (e.g., how much of the aperture is populated by dust grains), and the aperture size, $\rho$. $Af\rho$ should, in theory, be independent of aperture size, but we opt to use the 11\arcsec~radius aperture for reasons previously mentioned. Specifically, the equation for $Af\rho$ is as follows:

\begin{equation}
\label{Afp}
Af\rho ~=~\frac{\left (2r_{H}\Delta \right )^{2} }{\rho }\frac{F_{\rm comet}}{F_{\sun}}
\end{equation}

\noindent where $r_H$ is the heliocentric distance of the Sun in au, $\Delta$ is the geocentric distance of the comet in cm, $\rho$ is the projected aperture at the distance of the comet in arcseconds, and the $W2$ fluxes, $F_{\rm comet}$ and $F_{\sun}$, are the comet and solar fluxes, at 1 au, respectively. We adopt values from Jarrett et al. (2011) for the solar flux at $W1$, which are also mentioned in the Explanatory Supplement to the WISE Preliminary Data Release Products (Cutri et al. 2011). Using the $W1$ flux point, we create a model of the $W2$ flux and assume that the blackbody temperature of dust grains at this distance scales with heliocentric distance according to $T$\textsubscript{BB} = 286 $\times~r_{H}^{-1/2}$, adopted from Stevenson et al. (2015) and Rosser et al. (2018). The blackbody temperature at this distance is very low, so we then scale the $W1$ point to the solar spectrum and extrapolate the expected $W2$ contribution by creating a theoretical thermal curve with a Planck function for this blackbody temperature. This can be seen in Figure 2, showing the spectral energy distribution plots created for comet visits with obvious excess in the $W2$ signal thus indicating likely gaseous emission. The blue-dotted curve on the left-hand side is the reflected sunlight model, while the solid orange curve on the right-hand side is the theoretical thermal curve and the two curves are summed together with the dashed-black lines. The error bars refer to 1$\sigma$ measurements. In cases where there was a non-detection, data were ambiguous or poor in $W1$ from visual inspection of the images, we were only able to determine upper limits for $Af\rho$ by measuring the flux at the predicted position of the comet and make an estimation of flux from the remaining structure. A 3$\sigma$ upper limit on the flux gives a 3$\sigma$ upper limit on the production rate. However, in some cases there was remaining structure, as can be seen by visually inspecting the $W1$ and $W2$ images, where sometimes $W2$ appeared to be a point source. 

\subsection{CO and $CO_2$ Production Rates} \label{sec:dust}
Using the spectral energy distribution, the predicted $W2$ flux is determined from the $W1$ reflected component. Any excess signal in $W2$ after removing reflected and thermal components is assumed to be due to CO or $\rm CO_2$ gas emission. After determining if there is any excess signal in $W2$, we find the average column density along the line of sight to the comet assuming that it is due to all CO or all $\rm CO_2$ emission according to:

\begin{equation}
\label{Numcoldens}
\left \langle N \right \rangle~=~F_{\rm W_2} 4\pi \Delta ^{2}\frac{\lambda }{hc}\frac{r_{H}^{2}}{g}\frac{1}{\pi\rho ^{2}}
\end{equation}

\noindent adopted from Pittichov{\'a} et al. (2008), Stevenson et al. (2015), and Rosser et al. (2018). $\left \langle N \right \rangle$ is the average column density of the gas species along the line of sight in cm\textsuperscript{–2}, $F_{\rm W_2}$ is the excess flux in $W2$ after the dust contribution has been removed, $\lambda$ is the wavelength of observation in $\micron$, $h$ is Planck's constant in units of erg s, $c$ is the speed of light in a vacuum in units of cm s\textsuperscript{–1}, and $g$ is the fluorescence efficiency for a gas species that is taken to be ${2.86\times10^{-3}}$ s\textsuperscript{–1}, which is the fluorescence efficiency of ${\rm CO_2}$ at 1 au (Crovisier \& Encrenaz 1983), $r_H$ is the heliocentric distance of the Sun in au, $\Delta$ is the geocentric distance projected at the distance of the comet in cm, and $\rho$ is the projected aperture radius at the distance of the comet, respectively.\newline

Using this average column density, we determine the production rate due to $\rm CO_2$:

\begin{equation}
\label{Qprodrate}
Q_{\rm CO_2}~=~\left \langle N \right \rangle2\rho v \times10^5
\end{equation}

\noindent adopted from Pittichov{\'a} et al. (2008), Stevenson et al. (2015), and Rosser et al. (2018). $Q_{\rm CO_2}$ is the production rate of ${\rm CO_2}$ in molecules s\textsuperscript{–1}, $v$ is the expansion velocity of the gas species, which is adopted from the work of Ootsubo et al. (2012) who used 0.8 km s$\textsuperscript{–1} \times r_{H}^{-1/2}$, and $\times10^5$ is a unit conversion factor used previously in Stevenson et al. (2015) and Rosser et al. (2018) converting from km to cm, and $\rho$ is the scale in cm per arcsecond. Furthermore, because the ${\rm CO_2}$ emission line is a factor of $\sim$11.6 times stronger than that of CO, the equivalent production rates for CO (e.g., $Q_{\rm CO}$) can be found by multiplying $Q_{\rm CO_2}$ by 11.6 (Bauer et al. 2015; Rosser et al. 2018) because it would take as much CO to produce the same emission as ${\rm CO_2}$. This corresponds to the ratio of the fluorescence efficiencies for $\rm CO_2$ and CO. It should be noted that they cannot be separated because the $W2$ bandpass is so wide and because the fundamental CO $\nu$ = 1–0 and $\rm CO_2$ $\nu_3$ emission lines (4.67 $\micron$ and 4.23 $\micron$, respectively, for CO and $\rm CO_2$) both fall within the $W2$ channel. An assumption has to be made that the entire excess flux in this channel attributed to a gas species is due to solely CO or solely $\rm CO_2$. However, some comets have already been reported to have their gas flux component attributed mainly to the presence of CO rather than $\rm CO_2$, such as with 29P/Schwassmann-Wachmann 1 (Senay \& Jewitt 1994), C/1995 O1 (Hale–Bopp) (Crovisier et al. 1997; DiSanti et al. 2001; Biver et al. 2002; Gunnarsson et al. 2003), and C/2016 R2 (Pan-STARRS) (Wierzchos \& Womack 2018), helping to narrow down a closer value of the flux of the dominant constituent volatile species as the likely driver of cometary activity. For the cases of comets with measured $\rm CO_2$ production rates, measurements performed by Ootsubo et al. (2012) and McKay et al. (2019) have led to better estimates of $\rm CO_2$ inventories in comets. In any case, the triple-bonded molecular structure of CO and the fact that it has a longer photodissociation lifetime compared to $\rm CO_2$ (Huebner et al. 1992) are several key points to consider in the overall solar system volatile inventory, as well as for activity driven emission for comets at distances greater than $\gtrsim$4 au, where it is far too cold for water-ice sublimation to occur.\newline

Our measured $Q_{\rm CO_2}$ and $Q_{\rm CO}$ values are seen in Table 2 and are all between $\sim$10\textsuperscript{27} and $\sim$10\textsuperscript{29} molecules s\textsuperscript{–1}. For some cases where there were non-detections in $W1$ or poor quality data, such as B1's UT 2016 November visit, upper limits were made on $Af\rho$ thus setting lower limits on $Q_{\rm CO_2}$ as well as $Q_{\rm CO}$. However, if the actual $W1$ values are lower than reported, then our rates for $Q_{\rm CO_2}$ and $Q_{\rm CO}$ will be expected to be larger. It should be stated that the gas expansion velocity is unknown at this distance but is probably in the order of $\sim$0.2 km s$\textsuperscript{–1}$, especially for gaseous CO and $\rm CO_2$ species. This means that a difference in gas expansion velocity value would change the stated results.

\subsection{Simultaneous Observations of C/2017 K2 (Pan-STARRS) by the James Clerk Maxwell Telescope (JCMT) and the NEOWISE Satellite on UT 2021 April 1} \label{sec:simultaneousK2observations}
On UT 2021 April 1, Yang et al. (2021) took observations of K2 using the JCMT in Maunakea, Hawaii for an integration time of $\sim$8 hr. Similarly, on UT 2021 March 24, data were taken by the Hubble Space Telescope in the optical (HST GO Program, GO 16309; P.I. Jewitt). NEOWISE serendipitously scanned this region of the sky between UT 2021 March 30 to April 4 providing infrared data on morphology, as well as CO and $\rm CO_2$ production rates (this work) so as to not be encumbered by the Earth's $\rm CO_2$-rich atmosphere. In addition, similar morphological features were seen in both datasets. Our results show that we obtain CO and $\rm CO_2$ production rates of ${9.0\pm0.1\times10^{27}}$ mol s\textsuperscript{–1} for $Q_{\rm CO_2}$ and ${1.0\pm0.01\times10^{29}}$ mol s\textsuperscript{–1} for $Q_{\rm CO}$. Yang et al. (2021) reported $Q_{\rm CO}$ at a production rate of ${1.6\pm0.5\times10^{27}}$ mol s\textsuperscript{–1}. Although the NEOWISE data from $W2$  is potentially sensitive to both CO and $\rm CO_2$, the Yang et al. (2021) measurement is only of CO. Therefore, the simplest explanation is that the NEOWISE detection was of $\rm CO_2$ in comet K2, which was not seen in the JCMT data.\newline

The results of our study, Yang et al. (2021), and the work of Jewitt et al (2019b) show results over two regimes of the electromagnetic spectrum (e.g., optical and infrared). While not fully understood in terms of sublimation and outburst mechanisms, these results show that $Q_{\rm CO_2}$ and/or $Q_{\rm CO}$ activity does increase with decreasing heliocentric distance. Similar morphology can be seen in both works (see Figure 1 of this work and Figure 1 of Yang et al. 2021) but there appears to be an unusual trend of dust production rising and falling over the last three years for K2, as shown in Figure 3 of this work. We largely suspect that at the distance where activity for K2 first began to be noticed, there was ejection of dust driven by CO and/or $\rm CO_2$ sublimation. The results of Wierzchos \& Womack~(2020) showed that in the case of comet 29P, outbursting of CO did not always correlate with the dust outbursts. Because the surface properties of the comet are unknown, as well as its composition, this may have led to K2 switching between dust and then gas driven activity, until it moved closer inwards within the solar system where they both became dominant, as evident in the images seen in Figure 1, specifically, panels ((d.)–(l.)) showing its morphology with time and whether dust and/or gas were prevalent in the signal. Future studies may help to investigate this trend and also its significance with other comets. However, it may be such that either CO and/or $\rm CO_2$, or both were produced and appear to be the likely culprits driving activity on distant comets such as K2 and those arriving from the Oort cloud and the outermost regions of the solar system.\newline

\subsection{Evolution of Cometary Activity with $r_{H}$} \label{sec:cometaryevolution}
From our measurements, specifically in 2017 November where there is a detection in both $W1$ and $W2$, B1 appears to have an $Af\rho$ value of 1682 $\pm$ 119 cm just after perihelion passage. With regards to the measurements performed for B1 with an example in Figures 1 (a.)–(c.) of the B1 UT 2017 November visit. It is unknown how much of the remaining structure in the co-add may have affected the measured $Af\rho$ values. However, there is a much clearer detection in the $W2$ frames, which suggests that the detections may be real. As in previous studies of NEOWISE observed comets, detections more often occur in $W2$ compared to $W1$, which are likely to be an indicator of excess in gas flux driving activity.\newline

We note that for all 15 epochs of coverage for these comets, there appear to be detections in $W2$ whereas only nine of our comet visits have clear noticeable features indicating activity in $W1$. We generally assume that gas emission results in dust production, but gas emission flux would be drowned out if there is too much activity, especially if the dust is optically thick. This may be what is occurring in Figures 1 (d.)–(f.) (e.g., UT 2019 March–April) for K2. In addition, K2 is noteworthy in that our measured $Af\rho$ values are as low as 1853 $\pm$ 114 cm and as high as 6370 $\pm$ 30 cm. Jewitt et al. (2019b) reported that K2 would likely experience a noticeable change in activity some time in early-2019 as it passes around the region of $\sim$12 au, which is confirmed in our NEOWISE data and may be due to amorphous crystallization should amorphous ice exist on the comet. Amorphous crystallization refers to disordered structures, but with irregularity and the co-existence of both crystalline and amorphous water ice. These structures are also subject to irradiating sources such as the Sun and cosmic rays, which affect the organization of molecules, other volatiles, and the conditions of vapors at very cold temperatures at their initial formation (Jenniskens \& Blake, 1996). In the UT 2019 July visit, the excess $W2$ flux can be seen on the spectral energy distribution plot (see Figure 2(c.)) which corresponds to a $Q_{\rm CO}$ value of ${4.5\pm0.4\times10^{28}}$ mol s\textsuperscript{–1}. Also noteworthy is an evident trend as seen in Figure 4 showing that $Q_{\rm CO_2}$ measurements mostly increase with decreasing heliocentric distance. The gas production peaks between the UT 2020 March–April as well as UT 2020 August visits, and then falls during the UT 2021 March–April visit at about 6.72 au. However, with the $Af\rho$ measurements (e.g., dust production) specifically for K2, these seemingly rise during the satellite's second coverage in a given year. For example, for 2019, the first coverage is from March–April and the second is July–August. This trend has been obvious from 2019 to 2021, possibly suggesting an effect of solar radiation acting on some other volatile species present in the comet, which is worthy of future investigation.\newline

In the case of U3, it is interesting that most of its major activity (at least from visual observations; e.g., Hui et al. 2019) is suggested to have occurred much earlier than perihelion. Another result of this work is that U3 does not follow the expected convention of $Q_{\rm CO}$ and $Q_{\rm CO_2}$ increasing with decreasing distance to the Sun (unlike most of our results with B1 and K2). While U3 may be active in $W2$, it appears to be quiescing in activity in the later visits for gas production (see Figure 4). Because of (visual) non-detections in $W1$, we cannot comment on its dust production because we have only determined upper limits for $Af\rho$, thus setting lower limits for $Q_{\rm CO}$ and $Q_{\rm CO_2}$.\newline

Our work is unique in that we perform measurements of gas and dust for these three comets of interest over several visits during pre- and post-perihelion, with the latter specifically for B1. Figures 3 and 4 best summarize the activity of each comet from our measured $Af\rho$ and $Q_{\rm CO_2}$ values for each of the respective visits with heliocentric distance. In terms of $Q_{\rm CO_2}$ and/or $Q_{\rm CO}$ production with decreasing heliocentric distance, K2 mostly follows the expected convention of increasing with decreased heliocentric distance, U3's activity decreases with decreasing heliocentric distance, and B1 first increases with decreasing heliocentric distance, decreases, and then increases in the last visit. All comets have some coverage during the inbound legs of their orbits. This work has also covered the entire NEOWISE dataset while K2 was inbound from the years 2019–2022, first appearing as a point source and sometimes with extended features in the stacked images, and then in the individual frames around $\sim$8–9 au. Because K2 was still inbound (at the time of submission of this work), it will be interesting to see the long-term evolution of gas and dust production values, as well as the morphological aspects and what role, if any, $\rm CO_2$ and CO chemistry may have in what has been and what will be observed during its outbound orbit.\newline 

Previous studies have been reported for hyperactive comets near perihelion approach, such as 21P/Giacobini–Zinner (Pittichov{\'a} et al. 2008) and 46P/Wirtanen (Farnham et al. 2019). Few studies have been performed in a detailed manner over a long temporal baseline on the LPCs. However, Gunnarsson et al. (2003) performed a long-term study over 17 epochs on C/1995 O1 (Hale–Bopp), though all epochs of observation are when Hale–Bopp was outbound. It is our hope that this study will provide an informative and comparative benchmark for gas and dust production for LPCs active at large heliocentric distances (especially for the case of K2) in the years to come. This work, as well as future studies, may help us understand why certain LPCs become active at large distances from the Sun. Furthermore, NEOWISE is one of the last functioning observatories offering a unique capability for a large survey of cometary CO and $\rm CO_2$, until the advent of SPHEREx (Doré et al. 2018). However, the James Webb Space Telescope can also perform targeted studies on cometary morphology, as well as CO and $\rm CO_2$ production rates.


\section{Summary} \label{sec:summ}
Using NEOWISE reactivated mission data, we have performed a study of CO and $\rm CO_2$ production rates on remarkably active long-period comets B1, K2, and U3. This range of data over large heliocentric distances covers a large timespan during various portions of their orbits. The major results of this work are summarized as follows:\newline

\indent 1. Our most important result is that through NEOWISE photometry, we have determined $Q_{\rm CO_2}$ and $Q_{\rm CO}$ values between $\sim$10\textsuperscript{27} and $\sim$10\textsuperscript{29} molecules s\textsuperscript{–1}, respectively, attributed to excess flux in the $W2$ signal, for $\rm CO_2$ and CO species for some of the NEOWISE observations of B1 and K2. These values are not independent $Q_{\rm CO_2}$ and $Q_{\rm CO}$ measurements, and represent the amount of CO and $\rm CO_2$ being produced if one assumes that all of the gaseous emission is from CO (or $\rm CO_2$). Where excess flux in $W2$ exceeded that of the $W1$ signal, the corresponding results presented are our measured $Q_{\rm CO_2}$ and $Q_{\rm CO}$ production rate values. For the case of U3, we investigated if there was a purported outburst by using the NEOWISE data. As this comet has receded from the Sun, it appears to be declining in activity, as evident by its low production values. Though we have determined lower limits for the reported $Q_{\rm CO_2}$ and $Q_{\rm CO}$ rates for U3, they may still be comparable with B1 and K2.\newline

\indent 2. The range of $Af\rho$ dust production values for B1 (in all but one case) and K2 are all on the order of $\sim$10\textsuperscript{3} cm, suggestive of strong cometary hyperactivity. In the case of B1, peak dust emission takes place shortly after perihelion while K2 (now moving outbound) has strong dust-driven activity, even at heliocentric distances at or greater than $\sim$11 au. We correlate the rise in dust production with the onset of gaseous emission. It may be that at the onset of gas emission, dust particles may be ejected from the parent body, and there may be effects presently unknown to us (as the comet has not been visited in-situ). The fact also remains that the cometary composition of K2 is unknown, which may cause gas and/or dust production to become dominant over time. This is an interesting effect that we have noticed, but future studies will be needed on K2 to determine whether or not this may be significant. While U3 is a remarkable comet found to be active at a large heliocentric distance, its activity seems to have dropped off dramatically around $\sim$9 au, even as it was still approaching perihelion, possibly suggesting an almost full depletion of volatile materials, or some other unknown process. Because there are only three data points and this comet is faint and very far away, we can only surmise that the comet seems to be decreasing in CO and/or $\rm CO_2$ activity.\newline

\indent 3. From Figure 4, the three comets each have a distinct case of $Q_{\rm CO_2}$ and $Q_{\rm CO}$ production, sometimes following the expected convention of increased production rates with decreasing heliocentric distance when the comets are on the inbound leg of their orbits. Meanwhile, for the case of B1, we find that dust production was rising and peaked shortly after perihelion, and then declined with our $Af\rho$ measurements, whereas its gas production rates appear to rise in the last visit.\newline

We would like to acknowledge the assistance of D. Jewitt, R. M. Rich, E. Masongsong, P. Arriaga, A. Drozdov, F. Masci, R. Cutri, D. Kaisler, and E. L. Wright for helpful discussion, detailed comments, and reviews that contributed significantly to the improvement of this manuscript. We would also like to thank the two anonymous referees, the scientific editor, and the data editor with very helpful suggestions. This material is based upon work supported by the National Aeronautics and Space Administration under grant No. 80NSSC17K0407 issued through the NASA Education Minority University Research Education Project (MUREP) through the NASA Harriett G. Jenkins Graduate Fellowship activity. This publication makes use of data products from NEOWISE, which is a project of the University of Arizona and the Jet Propulsion Laboratory/California Institute of Technology, funded by the National Aeronautics and Space Administration, funded by the Planetary Science Division of the National Aeronautics and Space Administration. This research has made use of the NASA/IPAC Infrared Science Archive, which is funded by the National Aeronautics and Space Administration and operated by the California Institute of Technology. This research made use of Photutils, an Astropy package for detection and photometry of astronomical sources (Bradley et al. 2022).\newline

The Near-Earth Object Wide-field Infrared Survey Explorer Reactivation Mission (NEOWISE; Mainzer et al. 2014) is a NASA Planetary Science Division space-based survey to detect, track and characterize asteroids and comets, and to learn more about the population of near-Earth objects, including those that could pose an impact hazard to the Earth. NEOWISE systematically images the sky at 3.4 and 4.6 \micron, obtaining multiple independent observations on each location that enable detection of previously known and new solar system small bodies by virtue of the their motion. Because it is an infrared survey, NEOWISE detects asteroid thermal emission and is equally sensitive to high and low albedo objects.\newline

The Single-exposure Source Database is a compendium of position and flux information for source detections made on the individual NEOWISE 7.7 s $W1$ and $W2$ Single-exposure images. Because NEOWISE scanned the same region of the sky multiple times, the Single-exposure Database contains multiple, independent measurements of objects. Positions, magnitudes in the two NEOWISE bands, astrometric and photometric uncertainties, flags indicating measurement quality, the time of observations and associations with the AllWISE Source Catalog and 2MASS Point Source Catalog are presented for entries in the Database.\newline

This dataset or service is made available by the Infrared Science Archive (IRSA) at IPAC, which is operated by the California Institute of Technology under contract with the National Aeronautics and Space Administration.\newline

\facilities {NEOWISE, IRSA.} 
\software {Image Co-addition with Optional Resolution Enhancement (ICORE) (Masci 2013), Python, Astropy (Astropy Collaboration et al. 2013), Photutils (Bradley et al. 2022), SAOImage DS9 (Joye \& Mandel 2003), KaleidaGraph.}

\subsection{ORCID iDs} \label{sec:orcid}
\noindent Dave G. Milewski: 0000-0002-7841-9464\newline
\noindent Joseph R. Masiero: 0000-0003-2638-720X\newline
\noindent Jana Pittichová: 0000-0002-5736-1857\newline
\noindent Emily A. Kramer: 0000-0003-0457-2519\newline
\noindent Amy K. Mainzer: 0000-0002-7578-3885\newline
\noindent James M. Bauer: 0000-0001-9542-0953\newline



\begin{deluxetable*}{cclccrrccccccc}
\tabletypesize{\scriptsize}
\tablecaption{Orbital Elements for NEOWISE Observed Comets
\label{orbelem}}
\tablewidth{0pt}
\tablehead{\colhead{Object} & \colhead{UT Discovery Date} & \colhead{$a$\tablenotemark{a}} & \colhead{$e$\tablenotemark{b}} & \colhead{$i$\tablenotemark{c}} & \colhead{Epoch}  & \colhead{$q$\tablenotemark{d}} & \colhead{UT Perihelion Date}  
}
\startdata
C/2010 U3 & 2010 Oct 31 & –9881 & 1.00086 & 55.5 & 2000 & 8.446 & 2019 Feb 26  \\
C/2014 B1 & 2014 Jan 28 & –2973 & 1.00322 & 28.4 & 2000 & 9.557 & 2017 Sep 10  \\
C/2017 K2 & 2017 May 21 & –3830 & 1.00047 & 87.6 & 2000 & 1.798 & 2022 Dec 19  \\
\enddata

\tablenotetext{a}{Semimajor axis, in au.}
\tablenotetext{b}{Eccentricity.}
\tablenotetext{c}{Inclination, in degrees.}
\tablenotetext{d}{Perihelion distance, in au.}

\tablecomments{~All orbital elements taken from the JPL Small-Body Database Browser: https://ssd.jpl.nasa.gov/sbdb.cgi.}

\end{deluxetable*}

\clearpage

\begin{deluxetable*}{llcclllclcll}
\tabletypesize{\scriptsize}
\rotate
\tablecaption{Observation Geometry and Summary of Photometry for NEOWISE Observed Comets
\label{geometryphot}}
\tablewidth{0pt}
\tablehead{\colhead{Object} & \colhead{UT Date \tablenotemark{a}} & \colhead{$\#$ Frames} & \colhead{Detection?} & \colhead{$r_H$\tablenotemark{b}}  & \colhead{$\Delta$\tablenotemark{c}} & \colhead{$\alpha$\tablenotemark{d}} & \colhead{$F_{W_1}$\tablenotemark{e}} & \colhead{$F_{W_2}$\tablenotemark{f}} & \colhead{$Af\rho$\tablenotemark{g}} & \colhead{$Q_{\rm CO_2}$\tablenotemark{h}}  & \colhead{$Q_{\rm CO}$\tablenotemark{i}}  \\
\colhead{} & \colhead{} & \colhead{W1, W2} & \colhead{Yes or No} & \colhead{} & \colhead{} & \colhead{} & \colhead{} & \colhead{} & \colhead{} & \colhead{} & \colhead{} 
}
\startdata
C/2014 B1 & 2016 Nov 13 & 11, 11 & Y?, Y? & 9.70 $\downarrow$ & 9.66 & 5.9 & 0.018 & 0.344 & $\textless$277 & *${8.5\times10^{27}}$ & *${9.9\times10^{28}}$  \\
   & 2017 Apr 19 & 8, 9 & N, Y? & 9.59 $\downarrow$ & 9.24 & 5.7 & 0.149 & 0.532 & $\textless$2174 & ${*1.1\times10^{28}}$ & ${*1.2\times10^{29}}$  \\    
   & 2017 Nov 30 & 12, 12 & Y, Y & 9.57 $\uparrow$ & 9.52 & 5.9 & 0.112 $\pm$ 0.0081 & 0.231 $\pm$ 0.0153 & 1682 $\pm$ 119 & ${4.0\pm0.3\times10^{27}}$ & ${4.7\pm0.3\times10^{28}}$ \\  
     & 2018 May 2 & 11, 11 & N?, Y & 9.64 $\uparrow$ & 9.23 & 5.6 & 0.093 & 0.387 & $\textless$1654 & ${*8.0\times10^{27}}$ & ${*9.3\times10^{28}}$ \\
C/2017 K2 & 2019 Apr 1 & 30, 30 & Y, Y & 12.0 $\downarrow$ & 11.9 & 4.8 & 0.063 $\pm$ 0.0038 & 0.117 $\pm$ 0.0099 & 1853 $\pm$ 114 & ${3.4\pm0.3\times10^{27}}$ & ${4.0\pm0.3\times10^{28}}$  \\
& 2019 Jul 28 & 22, 22 & Y, Y? & 11.2 $\downarrow$ & 11.1 & 5.2 & 0.130 $\pm$ 0.0047 & 0.185 $\pm$ 0.0158 & 3105 $\pm$ 115 & ${3.9\pm0.3\times10^{27}}$ & ${4.5\pm0.4\times10^{28}}$  \\
& 2020 Mar 30 & 13, 13 & Y, Y & 9.49 $\downarrow$ & 9.45 & 6.0 & 0.144 $\pm$ 0.0060 & 0.326 $\pm$ 0.0172 & 2112 $\pm$ 90 & ${5.7\pm0.3\times10^{27}}$ & ${6.7\pm0.4\times10^{28}}$  \\
& 2020 Aug 4 & 16, 17 & Y, Y & 8.57 $\downarrow$ & 8.40 & 6.8 & 0.287 $\pm$ 0.0054 & 0.743 $\pm$ 0.0192 & 3042 $\pm$ 60 & ${1.1\pm0.03\times10^{28}}$ & ${1.2\pm0.03\times10^{29}}$  \\
& 2021 Apr 1 & 20, 20 & Y, Y & 6.72 $\downarrow$ & 6.65 & 8.6 & 0.799 $\pm$ 0.0063 & 1.370 $\pm$ 0.0165 & 4125 $\pm$ 35 & ${9.0\pm0.1\times10^{27}}$ & ${1.0\pm0.01\times10^{29}}$  \\
& 2021 Aug 1 & 11, 11 & Y, Y & 5.67 $\downarrow$ & 5.38 & 10.1 & 1.545 $\pm$ 0.0095 & 2.616 $\pm$ 0.0254 & 4595 $\pm$ 25 & ${1.1\pm0.01\times10^{28}}$ & ${1.2\pm0.01\times10^{29}}$  \\
& 2022 Apr 6 & 20, 20 & Y, Y & 3.52 $\downarrow$ & 3.38 & 16.5 & 5.233 $\pm$ 0.0567 & 12.269 $\pm$ 0.0827 & 3768 $\pm$ 40 & ${1.8\pm0.01\times10^{28}}$ & ${2.1\pm0.01\times10^{29}}$  \\
& 2022 Aug 5 & 8, 8 & Y, Y & 2.47 $\downarrow$ & 1.90 & 22.3 & 31.965 $\pm$ 0.1337 & 69.712 $\pm$ 0.2653 & 6370 $\pm$ 30 & ${3.2\pm0.01\times10^{28}}$ & ${3.8\pm0.01\times10^{29}}$  \\
C/2010 U3 & 2017 Sep 29 & 12, 12 & N, Y? & 8.97 $\downarrow$ & 8.92 & 6.4 & 0.052 & 0.277 & $\textless$637 & ${*5.2\times10^{27}}$ & ${*6.0\times10^{28}}$  \\
 & 2018 Feb 23 & 21, 21 & N, Y? & 8.72 $\downarrow$ & 8.43 & 6.3 & 0.034 & 0.219 & $\textless$369 & ${*3.8\times10^{27}}$ & ${*4.4\times10^{28}}$ \\ 
  & 2018 Nov 3 & 16, 16 & N, Y & 8.47 $\downarrow$ & 8.40 & 6.7 & 0.036 & 0.215 & $\textless$347 & ${*3.5\times10^{27}}$ & ${*4.1\times10^{28}}$  \\
\enddata

\tablenotetext{a}{UT date at midpoint of observation}
\tablenotetext{b}{Heliocentric distance, in au. $\downarrow$ or $\uparrow$ correspond to inbound or outbound orbits, respectively.}
\tablenotetext{c}{Geocentric distance, in au.}
\tablenotetext{d}{Phase angle, in degrees.}
\tablenotetext{e}{$W1$ flux, in millijanskies.}
\tablenotetext{f}{$W2$ flux, in millijanskies.}
\tablenotetext{g}{Dust production, in centimeters.}
\tablenotetext{h}{$\rm CO_2$ gas production, in molecules per second.}
\tablenotetext{i}{CO gas production, in molecules per second.}
\tablecomments{~In cases where there was an ambiguous or no clear detection of the object in $W1$, our listed $W1$ fluxes and $Af\rho$ values correspond to upper limits (denoted by $\textless$). $Q_{\rm CO_2}$ and $Q_{\rm CO}$~values with an (*) are calculated using the $W1$ $Af\rho$ upper limit. If the actual $W1$ values are lower than reported, then our rates for $Q_{\rm CO_2}$ and $Q_{\rm CO}$ will be expected to be larger. 
}
\end{deluxetable*}

\clearpage

\clearpage

\begin{figure*}
\figurenum{1}
\epsscale{0.85}
\plotone{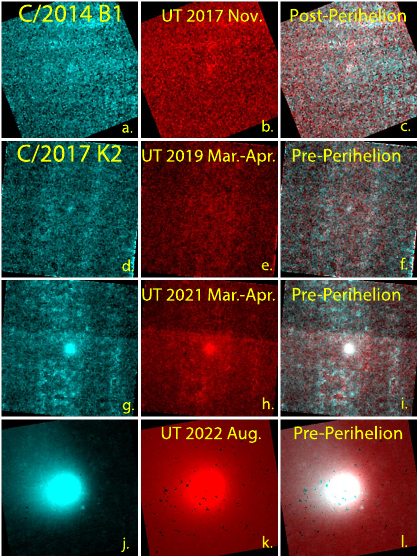}

\caption{Composite images made with ICORE (co-added in the moving reference frame of the comets and centered at its predicted ephemeris position from JPL HORIZONS) showing the comet at specific visits in $W1$ mapped to green and blue (left-hand panels), $W2$ mapped to red (center panels), and combined $W1$+$W2$ color-mapped composite frames (right-hand panels). For all images in Figure 1, the orientation (North is up, East is to the left) and image-size is consistent for all subsequent frames (e.g., $\sim$4.\arcmin5~$\times$~$\sim$4.\arcmin5) but the contrast and image stretch are not. Visible streaks are due to the slow motion of the comet, hence the background stars are not being fully suppressed in the stacking process. The name of the objects and dates of observations are labeled and applied for all panels in the rows. U3 has been excluded because its detection was ambiguous. See Table 2 for more information.
\label{fig1}}
\end{figure*}

\clearpage

\begin{figure}
\figurenum{2}
\epsscale{1.1}
\plotone{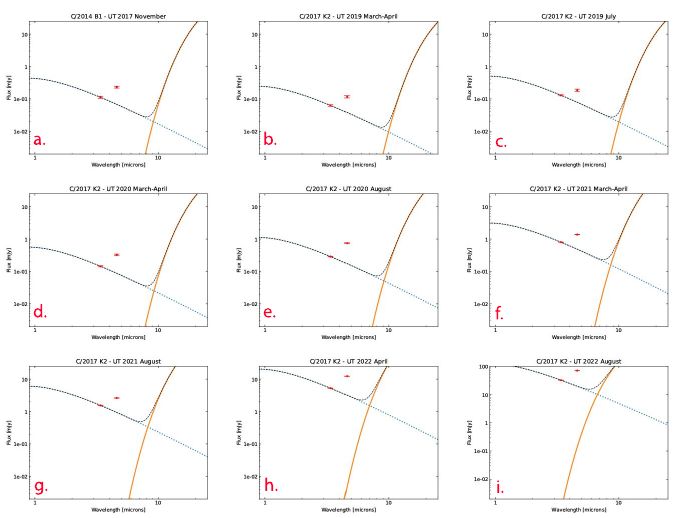}

\caption{Spectral energy distribution plots for comets where there is an excess in $W2$ flux which is above the $W1$ signal and assumed to be due to CO and/or $\rm CO_2$ gas emission. The corresponding names of the comets and visits are listed in labels. The left-hand curve (blue-dotted line) is a reflected sunlight model scaled to the 3.4 \micron~$W1$ point, while the theoretical thermal curve (orange solid line) is seen on the right-hand side. The two plots are summed together by a thin dashed-black line.
\label{fig2}}
\end{figure}

\clearpage

\begin{figure}
\figurenum{3}
\epsscale{1.2}
\plotone{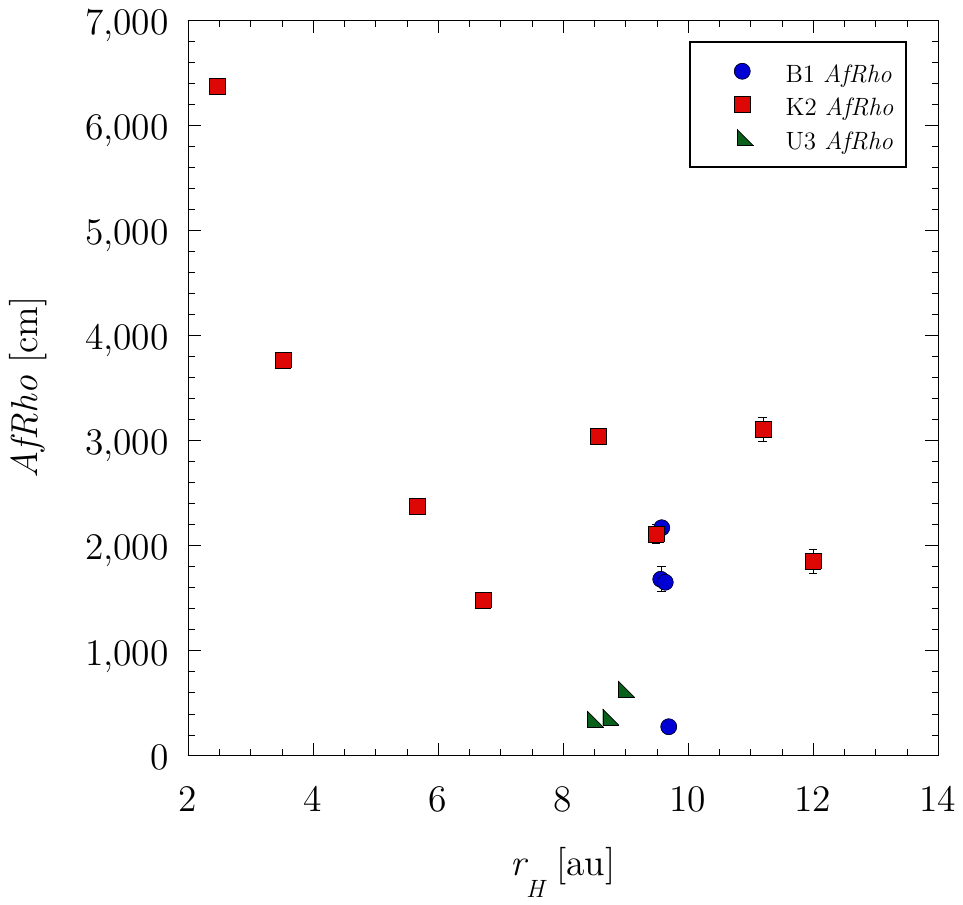}

\caption{Evolution of computed $Af\rho$ values seen over various NEOWISE visits as a function of heliocentric distance, $r_H$. B1, K2, and U3 $Af\rho$ values are denoted by blue circles, red squares, and green triangles for each respective comet. Note: because of some visits where there was an ambiguous or no clear $W1$ detection, these $Af\rho$ values correspond to upper limits.
\label{fig3}}
\end{figure}

\clearpage

\begin{figure}
\figurenum{4}
\epsscale{1.2}
\plotone{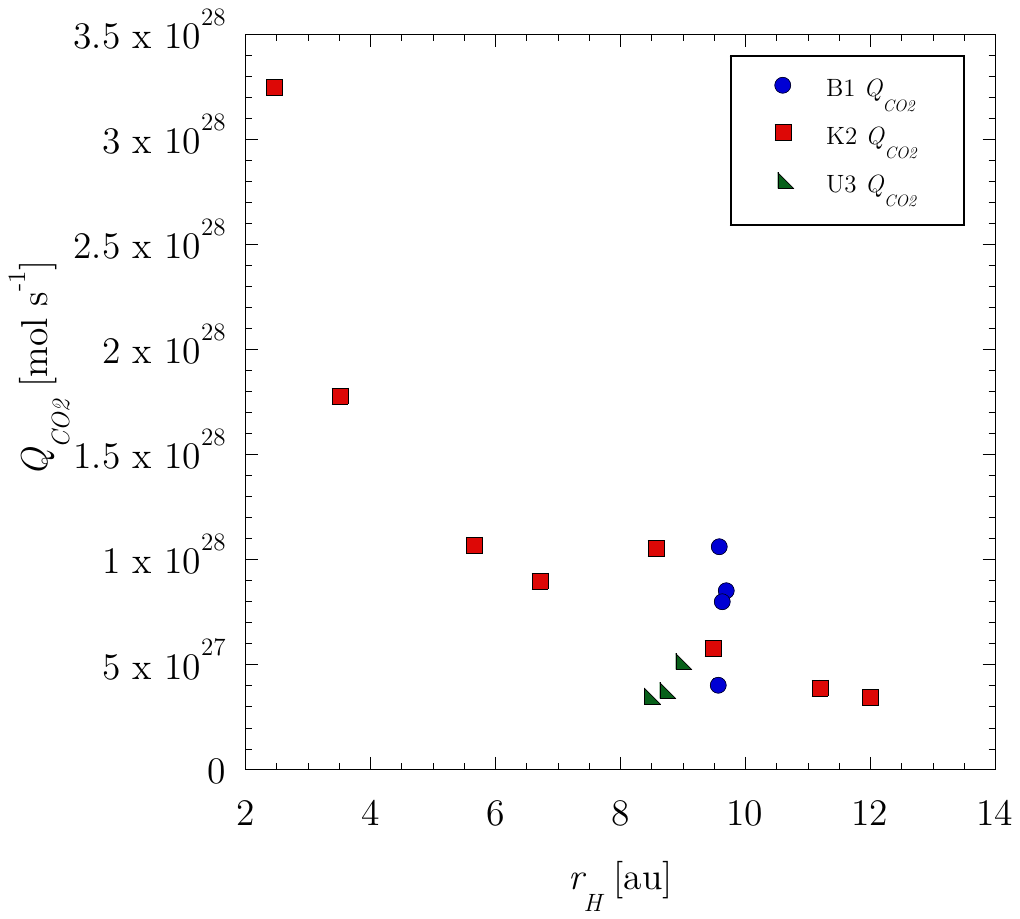}

\caption{Evolution of computed $Q_{\rm CO_2}$ values seen over various NEOWISE visits as a function of heliocentric distance, $r_H$. B1, K2, and U3 $Q_{\rm CO_2}$ values are denoted by blue circles, red squares, and green triangles for each respective comet. Note: because of some visits where there was an ambiguous or no clear $W1$ detection, the upper limit $Af\rho$ values determined in Figure 3 thus correspond to lower limits for $Q_{\rm CO_2}$ values.
\label{fig4}}
\end{figure}

\clearpage

\end{document}